\documentstyle[preprint,aps]{revtex}	
\input psfig
  
\def\zstar{Z^\star}
\def\wstar{W^\star}
\def\sighbar{\overline \sigma_{\h}}

\def\rts{\sqrt s}

\def\eg{{\it e.g.}}

\def\anti{\overline}

\def\mw{m_W}
\def\mz{m_Z}
\def\h{h}
\def\mh{m_{\h}}
\def\gamh{\Gamma_{\h}}
\def\hsm{h_{SM}}
\def\mhsm{m_{\hsm}}
\def\tanb{\tan\beta}
\def\hl{h^0}
\def\mhl{m_{\hl}}
\def\ha{A^0}
\def\mha{m_{\ha}}
\def\hh{H^0}
\def\mhh{m_{\hh}}
\def\fbi{~{\rm fb}^{-1}}

\def\gev{~{\rm GeV}}

\def\mt{m_t}

\def\overlay#1#2{\ifmmode \setbox 0=\hbox {$#1$}\setbox 1=\hbox to\wd 0{\hss
$#2$\hss }\else \setbox 0=\hbox {#1}\setbox 1=\hbox to\wd 0{\hss #2\hss }\fi
#1\hskip -\wd 0\box 1}
\def\case#1/#2{{\textstyle{#1\over#2}}}
\def\9{\phantom 0}      
\renewcommand\linebreak{\unskip\break} 
\def\lsim{\alt}

\catcode`@=11
\newcount\@tempcntc
\def\@citex[#1]#2{\if@filesw\immediate\write\@auxout{\string\citation{#2}}\fi
  \@tempcnta\z@\@tempcntb\m@ne\def\@citea{}\@cite{\@for\@citeb:=#2\do
    {\@ifundefined
       {b@\@citeb}{\@citeo\@tempcntb\m@ne\@citea\def\@citea{,}{\bf ?}\@warning
       {Citation `\@citeb' on page \thepage \space undefined}}%
    {\setbox\z@\hbox{\global\@tempcntc0\csname b@\@citeb\endcsname\relax}%
     \ifnum\@tempcntc=\z@ \@citeo\@tempcntb\m@ne
       \@citea\def\@citea{,}\hbox{\csname b@\@citeb\endcsname}%
     \else
      \advance\@tempcntb\@ne
      \ifnum\@tempcntb=\@tempcntc
      \else\advance\@tempcntb\m@ne\@citeo
      \@tempcnta\@tempcntc\@tempcntb\@tempcntc\fi\fi}}\@citeo}{#1}}
\def\@citeo{\ifnum\@tempcnta>\@tempcntb\else\@citea\def\@citea{,}%
  \ifnum\@tempcnta=\@tempcntb\the\@tempcnta\else
   {\advance\@tempcnta\@ne\ifnum\@tempcnta=\@tempcntb \else \def\@citea{--}\fi
    \advance\@tempcnta\m@ne\the\@tempcnta\@citea\the\@tempcntb}\fi\fi}
\catcode`@=12

\renewenvironment{thebibliography}[1]
 {\begin{list}{\arabic{enumi}.}
    {\usecounter{enumi} \setlength{\parsep}{0pt}
     \setlength{\itemsep}{3pt} \settowidth{\labelwidth}{#1.}
     \sloppy
    }}{\end{list}}

\def\mm{\mu^+\mu^-}
\def\ee{e^+e^-}

\def\tanb{\tan\beta}

\begin{document}

\newlength{\captsize} \let\captsize=\small 

\preprint{
\font\fortssbx=cmssbx10 scaled \magstep2
\hbox to \hsize{
$\vcenter{\hbox{\fortssbx University of California - Davis}
\hbox{\fortssbx University of Wisconsin - Madison}}$
\hfill$\vcenter{\hbox{\bf UCD-95-12} \hbox{\bf MADPH-95-884}
\hbox{\bf IUHET-299}
             \hbox{April 1995}}$ }
}

\title{\vspace*{1in}
$s$-Channel Higgs Boson Production \\ at a Muon-Muon Collider}

\author{V. Barger$^a$, M.S.~Berger$^b$, J.F.~Gunion$^c$, and T.~Han$^c$}
\maketitle
\begin{center}
\small\it
$^a$Physics Department, University of Wisconsin, Madison, WI 53706,
USA\\
$^b$Physics Department, Indiana University, Bloomington, IN 47405,
USA\\
$^c$Physics Department, University of California,  Davis, CA 95616, USA\\
\end{center}

\thispagestyle{empty}

\begin{abstract}

High luminosity muon-muon colliders would provide a powerful new probe of Higgs
boson physics through $s$-channel resonance production. We discuss the
prospects for detection of Higgs bosons and precision measurements of
their masses and widths at such a machine.
\end{abstract}

\newpage

The feasibility of constructing high luminosity muon-muon colliders is
currently under investigation \cite{sausi,sausii} and a first overview
of the phenomenology has been given \cite{workgr}.
The fact that the muon is 200
times more massive than the electron makes such colliders very attractive for
both practical and theoretical reasons:
\begin{description}
\item{(i)}  synchrotron radiation does not limit their circular acceleration
and
multi-TeV energies can be realized;

\item{(ii)} the beam energy resolution is not limited by beamstrahlung
smearing;

\item{(iii)} the $s$-channel production
of Higgs boson resonances ($\mu^+\mu^-\to \h$)
would make possible precision studies of the Higgs sector.

\end{description}

If electroweak symmetry breaking is realized via a scalar field Higgs sector,
then one of
the primary goals of future colliders must be to completely delineate the Higgs
spectrum and measure the Higgs masses, widths and couplings. In this Letter we
present a quantitative study of the merits of $s$-channel Higgs production at a
$\mu^+\mu^-$ collider with excellent beam energy resolution.

Two specific muon collider schemes are under consideration. A high energy
machine with 4~TeV center-of-mass energy ($\rts$) and luminosity of order
$10^{35}$~cm$^{-2}$~s$^{-1}$ \cite{palmer} would have an energy reach
appropriate for pair production of heavy supersymmetric particles \cite{workgr}
or, in the absence of Higgs bosons, the study of strong scattering of
longitudinally polarized $W$ bosons \cite{workgr,chano,vb}. A lower energy
machine, hereafter called the First Muon Collider (FMC), could have c.m.\
energy around 0.5~TeV with a luminosity of order $2\times
10^{33}$~cm$^{-2}$~s$^{-1}$ \cite{palmer} for unpolarized beams.
It is the latter machine that may be
most directly relevant to the $s$-channel Higgs process. The most costly
component of a muon collider is the muon source
(decays of pions produced by proton
collisions) and the muon storage rings would comprise a modest fraction of the
overall cost \cite{palmer2}. In order that full luminosity be maintained
at all c.m.\ energies where Higgs bosons are either observed or expected,
it is thus envisioned that multiple storage rings could
eventually be constructed with c.m.\ energies spanning the desired range.

For $s$-channel studies of narrow resonances, the energy resolution is an
important consideration. A Gaussian shape for the energy spectrum of each beam
is expected to be a good approximation, with an rms deviation
most naturally in the
range $R = 0.04$\% to 0.08\% \cite{jackson}. \ By additional cooling or
chromaticity corrections, this can either be decreased to $R = 0.01$\% or
increased to $R = 1$\%, respectively.
The corresponding rms error $\sigma$ in $\sqrt s$ is given by
\begin{equation}
\sigma = (0.04~{\rm GeV})\left({R\over 0.06\%}\right)\left({\sqrt s\over {\rm
100\ GeV}}\right) \ .
\label{resolution}
\end{equation}
The critical issue is how this resolution compares to the calculated total
widths of Higgs bosons. Widths for the Standard Model Higgs $\hsm$
and the three neutral Higgs bosons $\hl$, $\hh$, $\ha$
of the Minimal Supersymmetric Standard Model
(MSSM) are illustrated in Fig.~\ref{hwidthsprl}; for the MSSM Higgs bosons,
results at $\tan\beta = 5$ and 20 are shown.
For $R\alt 0.06\%$, the energy resolution in Eq.~(\ref{resolution})
can be smaller than the Higgs widths in many cases;
the total luminosity required for Higgs discovery by energy scanning
is also minimized by employing the smallest possible $R$.

The $s$-channel Higgs resonance cross section is
\begin{equation}
\sigma_{\h} = {4\pi\Gamma(\h\to\mu\mu) \Gamma(\h\to X)\over (s-\mh^2)^2 +
\mh^2\Gamma^2_{\h}} \ ,
\label{resonancexsec}
\end{equation}
where $\h$ denotes a generic Higgs boson which decays to a final state $X$. The
effective cross section $\sighbar$ is obtained by convoluting
with the Gaussian distribution in $\sqrt s$:
\begin{equation}
\sighbar = \int \sigma_{\h}(s')\ {\exp\left[-(\sqrt{s'}-\sqrt s)^2 \big/
(2\sigma^2)\right]\over \sqrt{2\pi} \sigma}\ d\sqrt{s'} \; .
\label{convolution}
\end{equation}
For $\sigma \gg \gamh$, $\sighbar$ at $\sqrt s = \mh$ is given by
\begin{equation}
\sighbar =  {\pi\gamh\over 2\sqrt{2\pi}\sigma}\ \sigma_{\h}(\sqrt s = m_{\h})
\label{narrowwidthxsec}
\end{equation}
and for $\gamh \gg \sigma$
\begin{equation}
\sighbar = \sigma_{\h}(\sqrt s = m_{\h}) \ .
\label{broadwidthxsec}
\end{equation}
Since the backgrounds vary slowly over the expected energy resolution
interval $\overline\sigma_B=\sigma_B$.
In terms of the integrated luminosity $L$, total
event rates are given by $L\overline\sigma$;
roughly $L=20\fbi$/yr is expected for the FMC.
Predictions for $\overline\sigma_{\hsm}$ for inclusive SM
Higgs production are given in Fig.~\protect\ref{hsmprl}
versus $\rts=\mhsm$ for resolutions of $R =
0.01$\%, 0.06\% and 0.1\%.
For comparison, the $\mm\to \zstar\to Z\hsm$ cross section
is also shown, evaluated at the value $\sqrt s = \mz + \sqrt 2 \mhsm$ for
which it is a maximum.

\goodbreak
\begin{flushleft}
{\bf SM Higgs Boson}
\end{flushleft}

The optimal strategy for SM Higgs {\it discovery} at a lepton collider is to
use the $\mm\to Z{\h}$ mode (or $\ee\to Z{\h})$ because no energy scan is
needed. Studies of $\ee$ collider capabilities indicate that the
SM Higgs can
be discovered if $\mhsm < 0.7 \sqrt s$ and its mass determined to a
precision \cite{barklow}
\begin{equation}
\Delta \mhsm \simeq 0.4\ {\rm GeV} \left({20\ {\rm fb}^{-1}\over
L}\right)^{{1\over 2}}
\label{massresolution}
\end{equation}
if $\mhsm \alt 140$~GeV. At the LHC the $\hsm \to \gamma\gamma$ mode is
deemed viable for $80 \alt \mhsm \alt 150$~GeV,
with a 1\% mass resolution \cite{cmsatlas}. Once
the $\hsm$ signal is found, the measurement of its width becomes the
paramount issue, and it is this task for which $s$-channel resonance production
at a $\mu^+\mu^-$ collider is uniquely suited.

For $\mhsm < 2\mw$ the dominant $\hsm$-decay channels
are $b\bar b$, $W\wstar$, and
$ZZ^\star$, where the star denotes a virtual weak boson. The light quark
backgrounds to the $b\bar b$ signal can be rejected by $b$-tagging.
For the $W\wstar$ and $Z\zstar$ channels we employ only
the mixed leptonic/hadronic modes ($\ell\nu2j$ for $W\wstar$ and $2\ell2j$,
$2\nu2j$ for $Z\zstar$, where $\ell = e$ or $\mu$ and $j$ denotes a quark jet),
and the visible purely-leptonic $Z\zstar$ modes
($4\ell$ and $2\ell2\nu$),
taking into account the major electroweak QCD backgrounds.
For all channels we assume a general signal and background
identification efficiency of $\epsilon= 50\%$, after selected acceptance
cuts \cite{further}. In the case of the $b\anti b$ channel, this is to
include the efficiency for tagging at least one $b$.
The signal and background channel cross sections
$\epsilon\overline\sigma BF(X)$ at $\rts=\mhsm$
for $X=b\anti b$, $W\wstar$ and $Z\zstar$
are presented in
Fig.~\ref{smratesprl} versus $\mhsm$ for a resolution $R = 0.06$\%;
$BF(X)$ includes the Higgs decay branching ratios for the signal,
and the branching ratios for the $W,\wstar$ and $Z,\zstar$
decays in the $Z\zstar$ and $W\wstar$ final states for both the signal and the
background.
Also shown for each channel is the statistical significance of the signal,
$n_\sigma = S/\sqrt B$, where $S$
and $B$ are the signal and background rates, $L\epsilon\overline\sigma BF(X)$;
an integrated luminosity of $L=1\fbi$ is assumed.
A detectable $s$-channel Higgs signal is
realizable for all $\mhsm$ values between the current LEP\,I limit and $2\mw$
except in the region of the $Z$~peak; a luminosity $L\sim 10\fbi$ at
$\sqrt s = \mhsm$ is needed for $85 \alt \mhsm \alt 100\gev$.

With $L=30\fbi$ devoted to an energy scan around $\rts=\mhsm$,
the total Higgs width could be determined with reasonable
accuracy. For example, for $\mhsm\sim 120\gev$ the error
$\Delta\Gamma_{\hsm}$ would be of order
$0.002\gev$ ($0.008\gev$) for $R=0.01\%$ ($R=0.06\%$).
In addition, the event rate in a given channel measures
$\Gamma({\hsm}\to\mu^+\mu^-)\times BF({\hsm}\to X)$.
Then, using the measured branching
fractions, the ${\hsm}\to\mu\mu$ partial width can be determined, providing an
important test of the Higgs coupling.

\begin{flushleft}
{\bf MSSM Higgs Bosons}
\end{flushleft}

The MSSM has three neutral Higgs bosons $\hl$ (CP-even), $\hh$
(CP-even), and $\ha$
(CP-odd). There is a theoretical upper bound on the mass of the lightest state
$\hl$ of $\mhl \alt 130$ to 150~GeV \cite{higgs1,higgs2}. If $\mha \agt 2\mz$
(typical of grand unified models), the couplings are approximately \cite{gh}
\begin{equation}
\begin{array}{cccc}
  &\mu^+\mu^-,b\bar b &t\bar t &ZZ,W^+W^- \\
\noalign{\vspace{4pt}}
\hl &1 &-1 &1 \\
\hh &\tan\beta &-1/\tan\beta &0 \\
\ha &-i\gamma_5\tan\beta &-i\gamma_5/\tan\beta &0
\end{array}
\label{couplings}
\end{equation}
times the SM-Higgs couplings. Thus the $\hl$ couplings are SM-like,
while $\hh,\ha$
have negligible $WW$ and $ZZ$ couplings and, for $\tanb>1$,
enhanced $\mm$ and $b\anti b$ couplings. If $\mha \alt \mz$,
then $\hh$
is SM-like, while $\hl$ has couplings like those for $\hh$ above.

A $\mm$ collider provides two particularly unique probes of the MSSM
Higgs sector.  First, the couplings of the SM-like MSSM Higgs boson deviate
sufficiently from exact SM Higgs couplings that it
may well be distinguishable from the $\hsm$ by measurements of
$\gamh$ and $\Gamma(\h\to\mm)$ at a $\mm$ collider, using the
$s$-channel resonance process. For instance, in the $b\anti b$ channel
$\gamh$ and $\Gamma(\h\to\mm)\times BF(\h\to b\anti b)$ can be measured
with sufficient accuracy so as to distinguish the $\hl$ from the $\hsm$
for $\mha$ values as high as $500\gev$
\cite{further}.

The second
dramatic advantage of a $\mm$ collider in MSSM Higgs physics is the ability
to study the non-SM-like Higgs bosons, \eg\
for $\mha\agt 2\mz$ the $\hh,\ha$. An $e^+e^-$ collider can only
study these states via $\zstar\to \ha\hh$ production, which could easily be
kinematically disallowed since GUT scenarios typically have $\mha\sim\mhh\agt
200$--250~GeV.  In $s$-channel production the $\hh,\ha$
can be even more easily observable than a SM-like Higgs.
This is because the partial widths
$\Gamma(\hh,\ha\to \mm)$ grow rapidly with increasing
$\tanb$, implying [see Eqs.~(\ref{narrowwidthxsec}) and (\ref{broadwidthxsec})]
that $\overline\sigma_{\hh,\ha}$ will become strongly enhanced
relative to SM-like values. $BF(\hh,\ha\to b\anti b)$ is also enhanced
at large $\tanb$,
implying an increasingly large rate in the $b\anti b$ final state.
Thus, we concentrate here on the $b\bar b$ final states of $\hh,\ha$
although the modes $\hh,\ha\to t\bar t$,
$\hh\to \hl\hl,\ha\ha$ and $\ha\to Z\hl$
can also be useful \cite{further}.

Despite the enhanced $b\anti b$ partial widths, the suppressed (absent)
coupling of the $\hh$ ($\ha$) to $WW$ and $ZZ$ means that,
unlike the SM Higgs boson, the $\hh$ and $\ha$ remain
relatively narrow at high mass, with widths $\Gamma_{\hh},\Gamma_{\ha}
\sim 0.1$ to 3~GeV. Since these widths are generally comparable to or
broader than the expected $\sqrt s$ resolution for $R = 0.06$\% and $\rts\agt
200\gev$,
measurements of these Higgs widths could be straightforward with a scan over
several $\sqrt s$ settings, provided that the signal rates are sufficiently
high. The results of a fine scan can be combined to get a coarse scan
appropriate for broader widths.

The cross section for $\mm\to \ha\to b\anti b$
production with $\tan\beta = 2$, $5$
and 20 (including an approximate cut and $b$-tagging efficiency of 50\%)
is shown versus $\mha$ in Fig.~\ref{susyprl}
for beam resolution $R = 0.06$\%. Also shown is
the significance of the $b\anti b$ signal for delivered luminosity $L =
0.1\fbi$ at $\sqrt s = \mha$. Discovery of the $\ha$ and $\hh$ will require an
energy scan if $\zstar\to \hh+\ha$
is kinematically forbidden; a luminosity of
20~fb$^{-1}$ would allow a scan over 200~GeV at intervals of 1~GeV with $L =
0.1$~fb$^{-1}$ per point. The $b\bar b$ mode would yield a $5\sigma$ signal at
$\sqrt s = \mha$ for $\tan\beta \agt 2$ for $\mha\lsim 2\mt$
and for $\tanb\agt5$ for all $\mha$.  For $\mha\agt \mz$ ($\mha\alt \mz$),
the $\hh$ ($\hl$)
has very similar couplings to those of the $\ha$ and would also
be observable in the $b\anti b$ mode down to similar $\tanb$ values.
Discovery of {\it both} the $\hh$ and $\ha$
MSSM Higgs bosons would be possible
over a large part of the $\mha\agt\mz$ MSSM parameter space.

In summary, $\mm$ colliders offer significant new opportunities for
probing the Higgs sector. The $s$-channel resonance production process is
especially valuable for precision Higgs mass measurements, Higgs width
measurements, and the search for Higgs bosons with negligible $\h ZZ$
couplings,
such as the $\hh,\ha$ Higgs bosons of the MSSM.
The techniques discussed here in the SM and MSSM theories
are generally applicable to searches for
any Higgs boson or other scalar particle that couples to $\mm$.

\section*{Acknowledgments}

This work was supported in part by the U.S.~Department of Energy under Grants
Nos.~DE-FG02-95ER40896, DE-FG03-91ER40674 and DE-FG02-91ER40661.
Further support was provided
by the Davis Institute for High Energy Physics
and by the University of Wisconsin Research
Committee, with funds granted by the Wisconsin Alumni Research Foundation.


\bigskip
\section*{Figures}

\begin{enumerate}

\item{
Total width versus mass of the SM and MSSM Higgs bosons, for $\tan\beta =
5$ and 20 in the MSSM.
\label{fig:fig1}}

\item{
Cross sections versus $\mhsm$ for inclusive SM Higgs production: (i) the
$s$-channel $\sighbar$ [Eq.~(\protect\ref{convolution})] for $\mm\to \hsm$
with $R = 0.01$\%, 0.06\% and 0.1\% and (ii) $\sigma(\mu^+\mu^-\to Z\hsm)$
at $ \protect\sqrt s = \mz + \protect\sqrt 2 \mhsm$.
\label{fig:fig3}}

\item{
The (a) $\hsm$ signal and (b) background
cross sections, $\epsilon \overline\sigma BF(X)$,
for $X=b\anti b$, and useful $W\wstar$ and $Z\zstar$ final states
(including a channel-isolation efficiency of $\epsilon=0.5$)
versus $\mhsm$ for SM Higgs $s$-channel production with
resolution $R=0.06\%$. Also shown: (c) the statistical significance
$S/\protect \sqrt B$ for the three channels for $L=1\fbi$.
\label{fig:fig4}}

\item{
(a) The effective $b\anti b$-channel cross section,
$\epsilon\overline\sigma_{\ha}BF(\ha\to b\bar b)$,
for $s$-channel production of the MSSM Higgs
boson $\ha$ versus $\protect\sqrt s=\mha$,
for $\tanb = 2$, 5 and 20, beam resolution $R = 0.06$\% and
channel isolation efficiency $\epsilon=0.5$; and
(b) corresponding statistical significance of the
$\ha\to b\anti b$ signal for $L=0.1\fbi$ delivered at
$\protect\sqrt s=\mha$.
\label{fig:fig5}}

\end{enumerate}

\begin{figure}[[ht]
\vskip 1in
\let\normalsize=\captsize   
\begin{center}
\smallskip
\begin{minipage}{12.5cm}       
\caption{
Total width versus mass of the SM and MSSM Higgs bosons, for $\tan\beta =
5$ and 20 in the MSSM.}
\label{hwidthsprl}
\end{minipage}
\end{center}
\end{figure}

\begin{figure}[[ht]
\let\normalsize=\captsize   
\begin{center}
\begin{minipage}{12.5cm}       
\smallskip
\caption{
Cross sections versus $\mhsm$ for inclusive SM Higgs production: (i) the
$s$-channel $\sighbar$ [Eq.~(\protect\ref{convolution})] for $\mm\to \hsm$
with $R = 0.01$\%, 0.06\% and 0.1\% and (ii) $\sigma(\mu^+\mu^-\to Z\hsm)$
at $\protect\sqrt s = \mz + \protect\sqrt 2 \mhsm$.}
\label{hsmprl}
\end{minipage}
\end{center}
\end{figure}

\begin{figure}[[ht]
\let\normalsize=\captsize   
\begin{center}
\bigskip
\begin{minipage}{12.5cm}       
\caption{
The (a) $\hsm$ signal and (b) background
cross sections, $\epsilon \overline\sigma BF(X)$,
for $X=b\anti b$, and useful $W\wstar$ and $Z\zstar$ final states
(including a channel-isolation efficiency of $\epsilon=0.5$)
versus $\mhsm$ for SM Higgs $s$-channel production with
resolution $R=0.06\%$. Also shown: (c) the statistical significance
$S/\protect \sqrt B $ for the three channels for $L=1\fbi$.}
\label{smratesprl}
\end{minipage}
\end{center}
\end{figure}

\begin{figure}[[ht]
\let\normalsize=\captsize   
\begin{center}
\begin{minipage}{12.5cm}       
\bigskip
\caption{
(a) The effective $b\anti b$-channel cross section,
$\epsilon\overline\sigma_{\ha}BF(\ha\to b\bar b)$,
for $s$-channel production of the MSSM Higgs boson
$\ha$ versus $\protect\sqrt s=\mha$,
for $\tanb = 2$, 5 and 20, beam resolution $R = 0.06$\% and
channel isolation efficiency $\epsilon=0.5$; and
(b) corresponding statistical significance of the
$\ha\to b\anti b$ signal for $L=0.1\fbi$ delivered at
$\protect\sqrt s=\mha$.}
\label{susyprl}
\end{minipage}
\end{center}
\end{figure}


\begin{thebibliography}{99}
\bibitem{further} Further details of our analysis will be provided in a
subsequent publication.
\end{thebibliography}
\end{document}